# Modern DDoS Attacks and Defences - Survey

Jonah Burgess, *Cyber-Security student, Queen's University Belfast*

*Abstract*—Denial of Service (DoS) and Distributed Denial of Service of Service (DDoS) attacks are commonly used to disrupt network services. Attack techniques are always improving and due to the structure of the internet and properties of network protocols it is difficult to keep detection and mitigation techniques up to date. A lot of research has been conducted in this area which has demonstrated the difficulty of preventing DDoS attacks altogether, therefore the primary aim of most research is to maximize quality of service (QoS) for legitimate users. This survey paper aims to provide a clear summary of DDoS attacks and focuses on some recently proposed techniques for defence. The research papers that are analysed in depth primarily focused on the use of virtual machines (VMs) (HoneyMesh) and network function virtualization (NFV) (VGuard and VFence).

## I. INTRODUCTION

Denial of service attacks are used to temporarily or permanently disrupt the availability of a computer, server or network so that legitimate users cannot access. Like many cyber-threats, DoS attacks have increased in prevalence and sophistication in recent years whilst decreasing in difficulty. The detection and mitigation techniques designed to defend against DDoS attacks struggle to keep up with the attacks and a lack of incentive on behalf of those who are not affected has reduced the potential for a distributed solution. There have been some high profile attacks on websites including Yahoo, CNN, Amazon and various government organizations around the world [8]. DDoS as a service or botnet rental has become very popular and has opened up these attacks to anybody who has the money and motivation to perform an attack, rather than just those who have the technical skills and knowledge.

### A. Attack Types and Techniques

The earliest DoS attacks were performed manually from a single computer system or network. The threat posed by these attacks severely increased when attackers began to acquire large networks of computers known as botnets and perform distributed denial of service attacks. These DDoS attacks rely on the attacker's ability to infect many systems which can be simultaneously instructed to attack a specific target through the use of a command and control application (C&C) which generally automates a lot of the process.

There are many specific attack types such as Smurf, TCP/SYN flood, UDP flood, Teardrop, Ping of Death, Land Attack, Ping Flood, Nuke attack etc. [2] These attacks can be broadly separated into two categories; Application and Network. The first attack category relies on vulnerabilities within a specific application being discovered and exploited whereas the latter focuses on vulnerabilities in how networks operate. Since networking protocols were initially designed without security in mind (particularly DoS in this case) and are yet to be replaced, they are inherently vulnerable to the attacks. Vulnerabilities in how networking equipment operates may also be exploited e.g. If a network is protected by a single firewall which has a vulnerability or misconfiguration, then it may only be necessary to disrupt the availability of this single device to block access to the entire network.

There are numerous categories of attack techniques which are used to perform DDoS attacks including bandwidth consumption, resource exhaustion and application exploitation [4]. Bandwidth consumption attacks involve flooding the network with more traffic than it is capable of handling. Resource exhaustion based attacks try to exhaust the target's resources (excluding bandwidth) and focus on the physical limitations or incorrect configuration of networking/server equipment. Application exploitation attacks take advantage of vulnerabilities and flaws in applications running on a network or server.

### B. Attacker Objectives and Motivation

There are several common types of attackers that can be broadly separated into the following categories. Professionals are usually botnet operators who offer DDoS as a service in exchange for payment or use their botnets to attack organizations until they agree to pay a ransom. Gamers commonly use DDoS either to ensure they win games or to frustrate and discredit specific opponents, these attackers generally have good technical knowledge and may work in groups e.g. DDoS against another team in a game. The motivation for gamers is generally to increase reputation rather than for financial gain. Opt-In attackers are a fairly recent category of users who agree to participate in a specific DDoS campaign (with the list of targets generally decided by a centralised C&C), this is generally an act of protest and the instigators make their motivation known the target and public in order to shine light on a specific issue. The objectives of the main groups of attackers listed above (and others not listed) fall into one of a few areas; extortion, espionage, protest and nuisance.

Submitted November 2016. This work was supported in part by Queens University, Belfast.

J. Burgess is with the Centre for Secure Information Technology (CSIT), Belfast, UK (e-mail: jburgess03@qub.ac.uk).

*C. Detection and Mitigation Techniques*

Many different techniques have been proposed for detecting and mitigating DDoS attacks, some general suggestions are provided in [4] and [8] gives a detailed comparison of some more specific approaches including IntServ, DiffServ, Class Based Queuing, Proactive Server Roaming, Resource Accounting, Resource Pricing, Pushback Approach and Throttling. Other common techniques include IP traceback, packet marking and filtering although these techniques have limitations. This paper does not cover many methods in great detail so for the main part of the survey we perform an in-depth analysis of techniques such as HoneyMesh, VGuard and VFence. These techniques are related in the sense that they all make use of virtual machines or virtual networking technologies to provide defence against DDoS attacks.

## II. HONEYMESH

HoneyMesh is a DDoS mitigation technique proposed in [2] which suggests the use of virtualized honeypots to prevent DDoS attacks and potentially identify attackers. The basic idea of using a honeypot on a network is to direct malicious traffic towards it whilst only allowing legitimate traffic to the actual server. To ensure that a honeypot appears like a real target to an attacker, the honeypot should run the same services and perform the same actions and as a genuine system. Another key feature of a honeypot is that it should monitor and log interactions with attackers, this allows for an investigation and potential prosecution.

Honeypots are categorized as high or low interaction based on whether they mimic all of the actions and services of a real system (high) or only those which attackers frequently scan or request (low). High interaction honeypots provide extra security but with additional resource and maintenance costs that aren't incurred by low interaction honeypots. Vulnerabilities are often left exposed on purpose to lure the attacker in, if they make a move they are more likely to trip an alert and the extra log information can be used for tracing the intruder.

Physical systems can be used as honeypots but it makes more sense to run them as virtual machines, this means that multiple honeypots can be used on a single physical server and if an attacker damages the honeypot it can easily be restored to its original state. Running honeypots on virtual machines is significantly cheaper than using real (physical) systems which has previously been proposed. Another previously proposed solution looks at using an Active Server in front of the server, this will analyse requests and forward legitimate traffic to the server but block any bad traffic. This solution was found to be effective but among other flaws, it negatively impacts the QoS of legitimate users because all requests must essentially pass through an extra gateway.

The new proposed solution suggests that a network of honeypot VMs could be placed onto a single physical server, known as a honeypot farm. It is possible to use a separate VM for each server type and which means each specific VM only runs the services that would be expected for that type of server. Once an attack has been detected on the network, all traffic from the attacker will be sent to the honeypot farm thereby ensuring that the real servers are accessible to legitimate users.

To ensure that actual servers are never subject to DDoS attacks, the solution also proposes having a pool of backup VMs which are normally left idle, they can then be activated if the attacker manages to compromise or crash a VM which is quite possible since there should be exposed vulnerabilities on the honeypots to attract attackers. A honeypot daemon (honey-d) should run on each physical server to provide an additional layer of protection in case the honeypot farm fails to detect an attack.

The honeypot VMs use machine learning to examine inbound traffic and determine whether the behaviour is malicious or benign. Initially each VM will need to be trained with a large amount of traffic to determine what is standard behaviour and what is malicious. If traffic is compared to the established baseline and appears to be malicious then the VM will verify this by sending a set of challenges to the potential attacker and waiting for a response. This action in itself slows down potential attackers, if the response to the challenge also appears to be suspicious then the VM sends more complex challenges, again slowing down the attacker. Once enough responses have been received to determine whether or not a DDoS attack is occurring, the VM can automatically take action to defend against the attack or allow the traffic through to the real server (if found to be legitimate). The honey daemons can be configured to provide the same functions, again providing another layer of defence in case the honeypot farm fails.

If a flooding attack is detected by a VM then the routing tables can be updated automatically to redirect all inbound traffic straight to the honeypot farm where the attacker will be faced with a series of challenges. This slows down the attacker and provides the time required to trace the origin whilst ensuring that the real servers are not affected by the attempted DDoS attack. The suspected attacker will only be required to respond to challenges whilst the honeypot farm is deciding whether the traffic is actually malicious and trying to track or at least identify them. If the user is confirmed to be malicious then all future traffic from that attacker can simply be blocked at the firewall.

Whilst flooding attempts are network based DDoS attacks, application based attacks that exploit vulnerabilities to crash a service and cause significant damage. These attacks are not as easy to detect and have more extreme consequences such as loss or theft of data and long term outages. The HoneyMesh overcomes this problem by ensuring that the VMs run the same services and have the same security features (or lack of) to lure attackers into a trap. The attacker may think that the DDoS attack has been a success but actually they are being provided challenges by the VM to track and block the malicious user. Since the DDoS attack has the potential to take down a VM and vulnerabilities that have been left open to lure attackers in could result in the a VM being compromised, it is important that a pool of backup VMs are waiting idle, ready to take over. Based on [2], Fig. 1 below demonstrates interaction between the HoneyMesh system and DDoS attackers and how the attacks are dealt with.

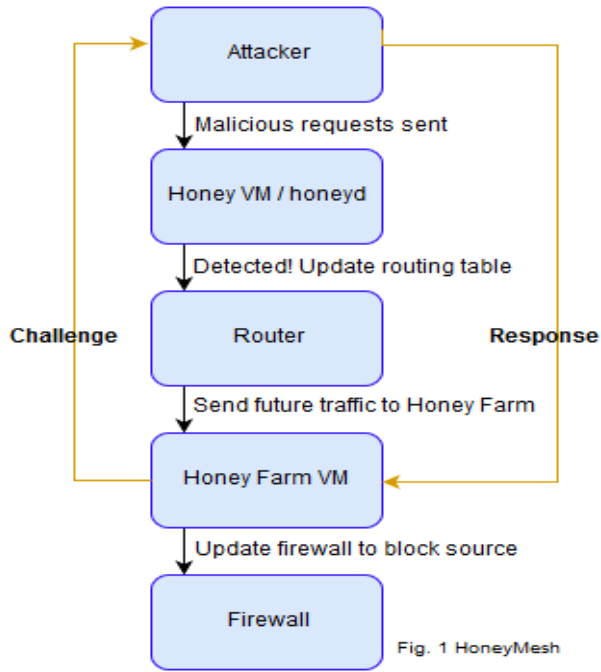

Fig. 1 HoneyMesh

There are some key advantages to the HoneyMesh solution that were found to be lacking in previous solutions. The honeypot farm deals with malicious traffic, preventing the real servers from being reached and ensuring the QoS is maintained for legitimate users. It is also very cost effective to run VMs as opposed to physical honeypots which had been proposed. Further, if a VM is compromised it is very easy to restore it and in the meantime there are backup VMs ready to take over.

One of the disadvantages of this solution is the proposal suggests that the entire network of VMs run from a single physical server, although this is more cost effective there is a risk that the entire physical server will be compromised or simply fail. To address this concern, the network of honeypots could be spread out between different physical servers.

## III. VGUARD

VGuard is a DDoS mitigation technique discussed in [5] which proposes that a virtual network function (VNF) be implemented in VMs to mitigate the risk of DDoS attacks. Network function virtualization (NFV) technology is currently still developing so this is a good time to look for potential solutions to mitigate DDoS attacks using this software-based networking technique. Due to the requirement of specialized hardware this solution would not have been realistic in the past but NFV allows VMs to run on basic hardware whilst performing specialized functions.

VGuard focuses specifically on attacks using real IP addresses rather than spoofed IP addresses as the latter is easier to mitigate and many effective solutions exist for this already. The principle function of the VGuard system is to separate malicious and legitimate traffic into different flows, whereby the legitimate traffic has a high priority and ensures a high QoS for users and the malicious traffic is labelled as low priority and users must compete for access to resources. The proposed solution outlines a static and dynamic method for NFV based DDoS detection which were both found to be effective.

The VGuard solution proposes that a firewall VNF and DDoS mitigation VNF be linked together to deal with inbound traffic. To prevent legitimate traffic being blocked due to being incorrectly labelled as malicious, the firewall VNF will only reject traffic which is guaranteed to be malicious (priority = 0). Similarly, traffic that is known to be benign (priority = 1) is passed into the DDoS mitigation VNF and forwarded directly into the high priority tunnel without delay.

When incoming traffic reaches the firewall and is not known to be unequivocally malicious or benign ($0 \leq$ priority $\leq 1$) it is forwarded to the DDoS mitigation VNF where it will be separated into the high or low priority tunnel based on how malicious it appears to be. If a benign user has a low priority level due to being incorrectly defined as malicious then their traffic will not be blocked, they will simply have to compete against other potentially malicious users so as not to impact connections coming from sources that seem legitimate. Fig. 2 taken from [5] shows architecture of the system.

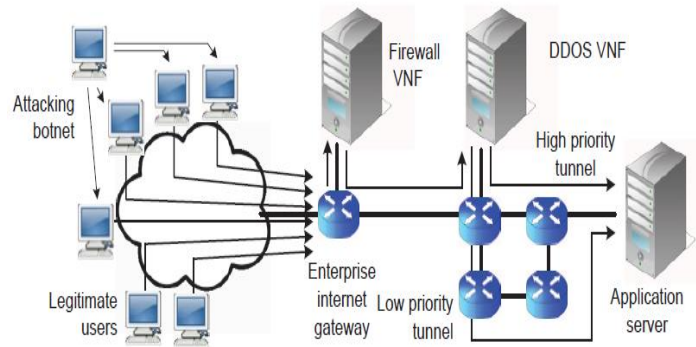

The DDoS mitigation VNF will regularly communicate with the server to ensure the log of source IPs and priority values are kept up to date. This means if a legitimate user has incorrectly been placed in the low priority tunnel but their actions are later proved to be benign then they can be reallocated into the high priority tunnel. Similarly, if a malicious user is misclassified as legitimate they will be redirected to the low priority tunnel after their actions have been detected. If malicious behaviour persists then eventually the priority value will reach 0 and access will be blocked entirely.

The combined traffic of the low priority and high priority tunnels must not exceed the available bandwidth. The proposed solution uses a utility analysis algorithm to determine the drop rate of both tunnels which is used to measure the satisfaction of users. The drop rate can in turn be used to determine the QoS provided to users so if no requests are dropped then the highest possible QoS has been provided. VGuard aims to maximize the overall QoS by applying both static and dynamic flow dispatching and comparing the effects.

The static flow dispatching algorithm uses flow priority distribution to try to ensure the optimal service is provided to users. If the data rate of all connections does not exceed the total bandwidth then it is possible to ensure balancing is applied so that neither tunnel's capacity is exceeded (therefore the QoS should be $= 1$, the maximum value). If either tunnel reaches its maximum capacity, then the threshold can be modified to ensure the greatest QoS.

The dynamic flow dispatching technique aims to overcome some of the limitations present in the static method. The key issues are that static flow dispatching relies on knowing the distribution of flow priority which is usually unknown and when the distribution changes, the optimal threshold may no longer be effective. To address this issue, the dynamic method is proposed which doesn't need to know the flow priority distribution.

The dynamic algorithm responds differently depending on the state of the tunnels. If neither tunnels have reached maximum capacity, then new flows will be assigned to the tunnel that has the most bandwidth available. When the high priority tunnel reaches a high capacity it will restrict future flow allocations so that only those with a high priority are allowed. If the high priority tunnel reaches the maximum limit of flows for the available bandwidth then all new flows are allocated to the low priority tunnel, regardless of their priority.

This solution was tested by implementing a Python-based simulator with two tunnels configured to use 50Mbps of bandwidth each and flows set to 100Kbps. Various scenarios were then tested by using different conditions such as flow inter-arrival times to determine how the system responded to different levels of inbound traffic.

Both the static and dynamic methods were tested against a DDoS attack which launched after 100 seconds and both were found to be very effective at ensuring the QoS for legitimate traffic maintained at a high level. The tests were run with VGuard activated and deactivated to confirm that the damage from the DDoS attack was severely reduced when the system was active. The tests also indicated that the static method was more beneficial in the steady state case whereas the dynamic technique was more effective when the flow distribution changes rapidly.

## IV. VFENCE

VFence is a DDoS mitigation technique discussed in [6] which uses the same NFV technology that is proposed in [5] (VGuard). Both proposals were released recently and feature some of the same authors but VFence is a newer proposal. Although both systems utilize the same technology, they have different aims. VGuard aims to mitigate DDoS attacks from real IP addresses which are controlled by a botnet but VFence looks at IP spoofing attacks which are used to mask the real address of the attacker whilst incurring far lower resource costs than real IP based DDoS attacks.

An example of a spoofed DDoS attack is SYN flood where the attacker sends a large number of TCP packets (using different spoofed source addresses) with the SYN flag set, initiating the three-way-handshake with the target server. If the victim's resources cannot cope with the level of SYN requests from the attacker, then legitimate users will be unable to initiate the handshake, rendering the service unusable. VFence specifically focuses on defence against SYN floods attacks.

The proposed solution uses a network of VNFs and multiple physical servers running virtual machines. The VMs are used to implement a dispatcher and multiple agents which sit between the client and the physical server, all inbound and outbound traffic flows through these VMs. When packets are sent to the server they first reach the dispatcher which forwards the packets to agents based on the source and destination address of the packets. The dispatcher keeps a log of this information and notes which agents the packets have been assigned to, this allows for load balancing between the agents.

The agents perform the actual filtering of traffic using a whitelist and filtering rules but also have another key function specifically related to the SYN flood attack. This function is a spoofed handshaking process which allows the agents to verify the legitimacy of the source address of the packet. In order to verify the source, the agent uses a SYN cookie to generate a SYN-ACK packet and respond to the original SYN packet. The SYN cookie sets a sequence number which is created using the current time along with source and destination address of the SYN packet received from the client.

If the client receives the spoofed SYN-ACK packet (the source was spoofed after all) they return an ACK packet which is forwarded back to the agent by the dispatcher to verify its legitimacy. This verification involves comparing the ACK number with the sequence number created using the SYN cookie. If the source is found to be legitimate then the source address is whitelisted and the agent initiates a spoofed three-way-handshake with the physical server. Once this spoofed handshake is completed, the connection between the client and server is opened and future communications will be forwarded between the client and server (by the same agent) without any extra verification.

Fig. 3a taken from [6] demonstrates a successful connection from a legitimate client whilst Fig. 3b shows the flow pattern of a failed connection from a malicious client. Note that in Fig. 3a there is an extra delay which is the result of special case that occurs when there is a delay between the time that a client's source address is verified and the time it is added and the time that the handshake with the server is completed or the source is added to the whitelist. When this occurs the agent will hold on to the data packets received from the client until the connection is fully established and verified.

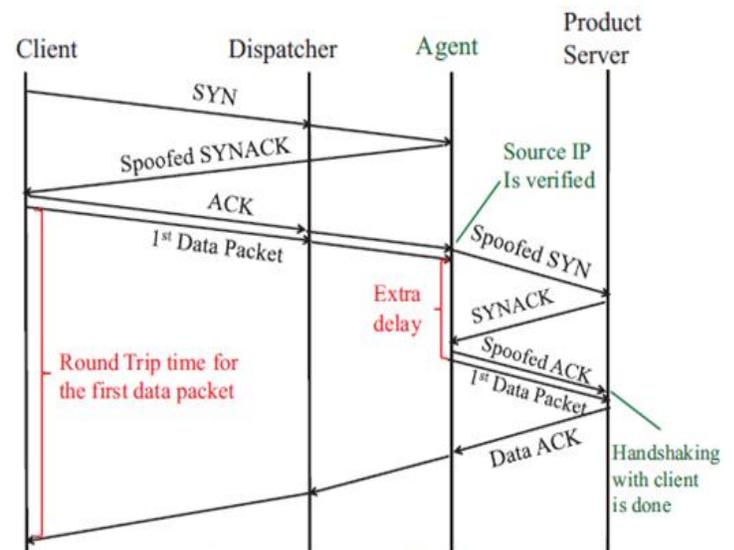

Fig. 3 a Successful connection

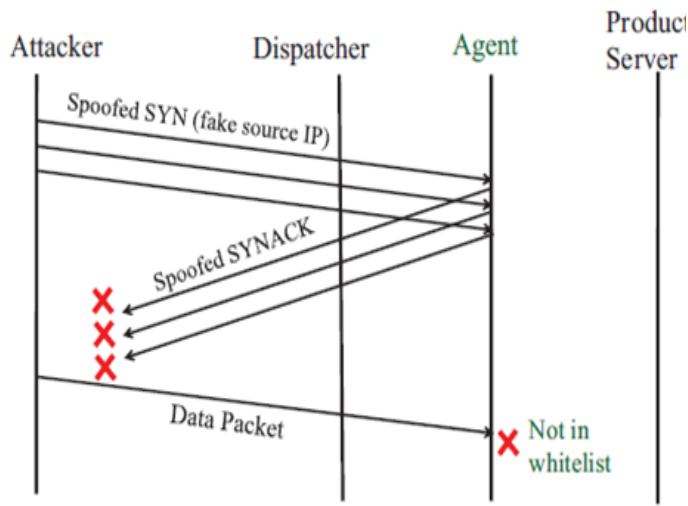

Fig 3 b. Rejected connection

If the attacker is using a spoofed address to send the SYN packet, then they would not receive the SYN-ACK packet from the agent. If they try to spoof an ACK packet, then the agent will also reject the request because the sequence number would not match up with the one generated by the SYN cookie earlier. The proposed solution suggests that the agents should be deployed dynamically so that the number of agents available to deal with requests increases or decreases depending on the demand at the time. How this can be achieved is discussed briefly but a more detailed solution is left for future work.

Like the VGuard solution, the VFence solution is tested through simulations. A Java program was developed (rather than Python) to accomplish this and the simulation network had 2 clients, a dispatcher, 5 agents, a switch, and a server. The server can process 200 pkts/sec and each agent can process 1000 pkts/sec. One of the clients sends normal traffic (100 pkts/sec) and the other sends malicious traffic (beginning 10 seconds in) by means of a SYN flood which gradually increases in intensity until it reaches 1000 pkts/sec (15 seconds in) and remains at this level until the experiment ends (30 seconds). Fig. 4 taken from [6] shows the results of this test.

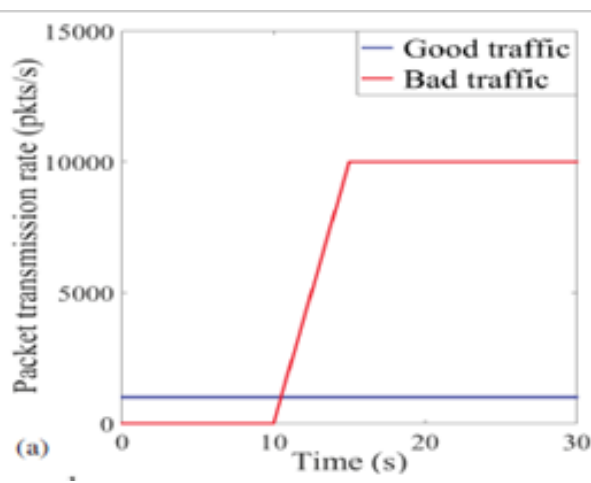

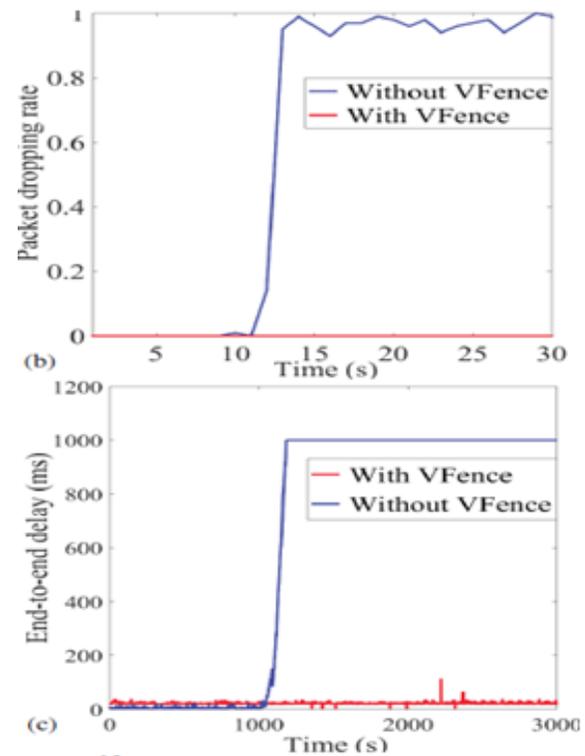

(a) shows the rate of packet transmission throughout the experiment. (b) shows the rate that packets are dropped at and we can see that no packets are lost when VFence is active. (c) shows the performance of the system (QoS) and we can see this is fairly stable when VFence is active but there is a huge delay if VFence is not used. Overall the results from this experiment indicate that this solution is very effective at defending the network against the SYN flood attack.

V. CONCLUSION

This survey introduced some basic DDoS concepts, attack types, attacker motivations/objectives before analysing some potential DDoS mitigation techniques in depth. The solutions discussed are based on recent, cutting edge research in the area of virtual machines and virtual networking. The three specific mitigation techniques studied were HoneyMesh (farm of honeypot VMs), VGuard and VFence (NFV based solutions). HoneyMesh proposed a solution which has many logical advantages over existing solutions and covered both network and application based DDoS attacks. Unfortunately, the technique was not tested in a real environment so the effectiveness could not be accurately measured.

VGuard and VFence both used the same NFV technology to implement the solutions although they addressed different categories of DDoS attacks with very different methods. VGuard addressed DDoS attacks that use real IP addresses which occur when a botnet is used whereas VFence defended against attacks that use spoofed IP addresses, the SYN flood was specifically tested for this purpose. These proposed solutions were evaluated using experimental simulations.

Since all of these solutions were proposed very recently future work should aim to test these techniques in real environments to help determine any improvements which could be applied and how other attacks can be prevented.